# An experimental verification of the one-dimensional static Willis-form equations

R.W. Yao[‡], H.X. Gao[‡], Y.X. Sun, X.D. Yuan, Z.H. Xiang[*]

*AML, Department of Engineering Mechanics, Tsinghua University, Beijing 100084, China*

**ABSTRACT**

This paper investigates the behavior of a heavy soft spring in steady circular motion. Since the spring is inhomogeneous due to centrifugal force, one can rigorously prove that it follows the one-dimensional static Willis-form equations. The theoretical predictions agree very well with experimental results. It further demonstrates that these equations can give a clear understanding of the stress-stiffening and spin-softening effect. These findings reveal that the Willis-form equations can give very accurate linear approximations of finite deformation problems and are also helpful to clarify the classical concept of the principle of material frame indifference.

## 1. Introduction

About three decades ago, Professor Willis (1981, 1997) proposed new linear elastic equations for inhomogeneous media by using the perturbation theory. The Willis equations include a constitutive equation

[*] Corresponding author. *Email address*: xiangzhihai@tsinghua.edu.cn.

[‡] These authors contributed equally to this paper.

$$\langle \boldsymbol{\sigma} \rangle = \boldsymbol{C}_{eff} \langle \boldsymbol{e} \rangle + \boldsymbol{S}_{eff} \langle \dot{\boldsymbol{u}} \rangle \tag{1a}$$

and an equation of motion

$$\nabla \cdot \langle \boldsymbol{\sigma} \rangle + \boldsymbol{f} = \bar{\boldsymbol{S}}_{eff} \langle \dot{\boldsymbol{e}} \rangle + \boldsymbol{\rho}_{eff} \langle \ddot{\boldsymbol{u}} \rangle , \tag{1b}$$

where $\boldsymbol{\sigma}$ is the stress tensor; $\boldsymbol{u}$ is the displacement vector free of rigid translations; $\boldsymbol{e}$ is the strain tensor, $\boldsymbol{e} = (\nabla \boldsymbol{u} + \boldsymbol{u}\nabla)/2$; $\boldsymbol{f}$ is the body force vector; $\langle \ \rangle$ denotes the ensemble average; the superposed dot denotes the differentiation with respect to time $t$; $\boldsymbol{C}_{eff}$, $\boldsymbol{S}_{eff}$, $\bar{\boldsymbol{S}}_{eff}$ and $\boldsymbol{\rho}_{eff}$ are non-local operators acting on $\langle \boldsymbol{e} \rangle$, $\langle \dot{\boldsymbol{u}} \rangle$, $\langle \dot{\boldsymbol{e}} \rangle$ and $\langle \ddot{\boldsymbol{u}} \rangle$ respectively. According to Willis (1997), the details of these operators are:

$$\boldsymbol{C}_{eff} \langle \boldsymbol{e} \rangle = \left( \boldsymbol{C}^0 + \langle \delta \boldsymbol{C} \rangle \right) : \langle \boldsymbol{e} \rangle - \left\langle \left( \delta \boldsymbol{C} - \langle \delta \boldsymbol{C} \rangle \right) : \boldsymbol{S}_x : \left( \delta \boldsymbol{C} - \langle \delta \boldsymbol{C} \rangle \right) \right\rangle \circ \langle \boldsymbol{e} \rangle , \tag{2a}$$

$$\boldsymbol{S}_{eff} \langle \dot{\boldsymbol{u}} \rangle = - \left\langle \left( \delta\rho - \langle \delta\rho \rangle \right) \left( \delta \boldsymbol{C} - \langle \delta \boldsymbol{C} \rangle \right) : \boldsymbol{M}_x \right\rangle \circ \langle \dot{\boldsymbol{u}} \rangle , \tag{2b}$$

$$\bar{\boldsymbol{S}}_{eff} \langle \dot{\boldsymbol{e}} \rangle = - \left\langle \left( \delta\rho - \langle \delta\rho \rangle \right) \boldsymbol{S}_t : \left( \delta \boldsymbol{C} - \langle \delta \boldsymbol{C} \rangle \right) \right\rangle \circ \langle \dot{\boldsymbol{e}} \rangle , \tag{2c}$$

$$\boldsymbol{\rho}_{eff} \langle \ddot{\boldsymbol{u}} \rangle = \left( \rho^0 + \langle \delta\rho \rangle \right) \langle \ddot{\boldsymbol{u}} \rangle - \left\langle \left( \delta\rho - \langle \delta\rho \rangle \right) \boldsymbol{M}_t \left( \delta\rho - \langle \delta\rho \rangle \right) \right\rangle \circ \langle \ddot{\boldsymbol{u}} \rangle , \tag{2d}$$

where $\boldsymbol{C}^0$ and $\rho^0$ are constant elasticity tensor and density, respectively; $\delta \boldsymbol{C}$ and $\delta\rho$ are corresponding small perturbations; $\boldsymbol{S}_x$, $\boldsymbol{M}_x$, $\boldsymbol{S}_t$ and $\boldsymbol{M}_t$ are tensors related to the Green's function $\boldsymbol{G}(\boldsymbol{x}, \boldsymbol{x}', t)$ with the source point $\boldsymbol{x}'$ and the field point $\boldsymbol{x}$; and $\circ$ is an operator representing field integration and time-convolution. A detailed example of $\circ$ is

$$\begin{aligned} & \left\langle \left( \delta\rho - \langle \delta\rho \rangle \right) \boldsymbol{M}_t \left( \delta\rho - \langle \delta\rho \rangle \right) \right\rangle \circ \langle \ddot{\boldsymbol{u}} \rangle \\ & \to \int_\Omega \left[ \left\langle \left( \delta\rho - \langle \delta\rho \rangle \right)(\boldsymbol{x}') (M_t)_{rs} (\boldsymbol{x}', \boldsymbol{x}, t) \left( \delta\rho - \langle \delta\rho \rangle \right)(\boldsymbol{x}) \right\rangle * \langle \ddot{u}_s \rangle (\boldsymbol{x}, t) \right] \mathrm{d}\boldsymbol{x} \end{aligned} , \tag{3}$$

where $*$ denotes the time-convolution.

In contrast to the classical linear elastic equations, the Willis equations are more complex and abstract. Therefore, not much attention had been paid to them until the upsurge of designing metamaterials with transformation methods (Shamonina and Solymar, 2007).

In the transformation method, a wave equation in a virtual space is transformed into a physical space through local mapping functions of independent variables (coordinates) and dependent variables (field variables). If this equation is form-invariant after the transformation, one can obtain the effective material properties by comparing the corresponding terms in the original and transformed equations. The resultant material in the physical space is called a metamaterial that is inhomogeneous and can steer the wave front along desired trajectories. Although the transformation method had achieved great successes in optics (Leonhardt, 2006; Pendry et al., 2006) and acoustics (Chen and Chan, 2010), it met much difficulty in elastics, because Milton et al. (2006) found that upon applying the local transformation on the classical linear elastic equations in frequency domain results in the corresponding Willis equations in local form. In another word, the classical linear elastic equations are not form-invariant under the local transformation but the Willis equations do.

The findings of Milton et al. (2006) aroused much interest in studying the Willis equations. Most of these studies focused on the dynamic homogenization of periodically inhomogeneous media (Willis, 2011; Norris et al., 2012; Srivastava and Nemat-Nasser, 2012; Nassar et al., 2015, 2016; Torrent et al., 2015; Srivastava, 2015), which involve non-local properties. However, if using transformation methods with point to point mappings, one can naturally obtain local properties (Milton et al., 2006).

For example, Xiang (2014) proved that the form-invariance is an intrinsic character of wave equations independent of the gauge adopted in the linear local transformation. He also obtained the Willis equations in time domain by using the deformation gradient as the transformation gauge for displacements:

$$\boldsymbol{\sigma} = \boldsymbol{C} : \boldsymbol{e} + \boldsymbol{S} \cdot \boldsymbol{u}, \tag{4a}$$

$$\nabla \cdot \boldsymbol{\sigma} + \boldsymbol{f}^a = \boldsymbol{S}^{\mathrm{T}} : \boldsymbol{e} + \boldsymbol{K} \cdot \boldsymbol{u} + \boldsymbol{\rho} \cdot \ddot{\boldsymbol{u}}, \tag{4b}$$

where $\boldsymbol{C}$ is the symmetric elasticity tensor; $\boldsymbol{\rho}$ is the mass density tensor; and $\boldsymbol{f}^a$ is an additional external body force in the physical space; $\boldsymbol{S}^{\mathrm{T}} : \boldsymbol{e}$ means $S_{ijk} e_{ij}$.

By noticing the differential property of time convolution

$$\frac{\mathrm{d}}{\mathrm{d}t}(\varphi * \psi) = \dot{\varphi} * \psi = \varphi * \dot{\psi}, \tag{5}$$

one can easily prove that the operator $\circ$, as shown in Eq. (3), has the similar property:

$$\frac{\mathrm{d}}{\mathrm{d}t}(\boldsymbol{\varphi} \circ \boldsymbol{\psi}) = \dot{\boldsymbol{\varphi}} \circ \boldsymbol{\psi} = \boldsymbol{\varphi} \circ \dot{\boldsymbol{\psi}}. \tag{6}$$

Thus, Eq. (1) can also be written as:

$$\langle \boldsymbol{\sigma} \rangle = \boldsymbol{C}_{eff} \langle \boldsymbol{e} \rangle + \boldsymbol{D}_{eff} \circ \langle \boldsymbol{u} \rangle, \tag{7a}$$

$$\nabla \cdot \langle \boldsymbol{\sigma} \rangle + \boldsymbol{f} = \bar{\boldsymbol{D}}_{eff} \circ \langle \boldsymbol{e} \rangle + \boldsymbol{K}_{eff} \circ \langle \boldsymbol{u} \rangle + \rho \langle \ddot{\boldsymbol{u}} \rangle, \tag{7b}$$

where

$$\boldsymbol{D}_{eff} = -\left\langle \left(\delta\rho - \langle \delta\rho \rangle\right)\left(\delta\boldsymbol{C} - \langle \delta\boldsymbol{C} \rangle\right) : \dot{\boldsymbol{M}}_x \right\rangle, \tag{8a}$$

$$\bar{\boldsymbol{D}}_{eff} = -\left\langle \left(\delta\rho - \langle \delta\rho \rangle\right) \dot{\boldsymbol{S}}_t : \left(\delta\boldsymbol{C} - \langle \delta\boldsymbol{C} \rangle\right) \right\rangle, \tag{8b}$$

$$\boldsymbol{K}_{eff} = -\left\langle \left(\delta\rho - \langle \delta\rho \rangle\right) \ddot{\boldsymbol{M}}_t \left(\delta\rho - \langle \delta\rho \rangle\right) \right\rangle, \tag{8c}$$

$$\rho = \rho^0 + \langle \delta\rho \rangle. \tag{8d}$$

Eq. (7) is in the same form as that of Eq. (4), which implies that Eq. (4) is also a possible result under the original theoretical framework of Willis (1997).

With the presence of pre-stresses $\boldsymbol{\sigma}^0$, Xiang and Yao (2016) further proved that the $\boldsymbol{S}$ in Eq. (4) is the gradient of $\boldsymbol{\sigma}^0$ and $-\boldsymbol{K}\cdot\boldsymbol{u}$ equals to the increment of the body force due to the change of the small perturbation from the pre-stressed configuration to the current configuration. They also pointed out that these equations are still valid for static problems if the inertia is ignored, namely,

$$\boldsymbol{\sigma} = \boldsymbol{C} : \boldsymbol{e} + \left(\boldsymbol{\sigma}^0\nabla\right)\cdot\boldsymbol{u}, \tag{9a}$$

$$\nabla\cdot\boldsymbol{\sigma} + \boldsymbol{f}^a = \left(\boldsymbol{\sigma}^0\nabla\right)^{\mathrm{T}} : \boldsymbol{e} + \boldsymbol{K}\cdot\boldsymbol{u}, \tag{9b}$$

where $\left(\boldsymbol{\sigma}^0\nabla\right)^{\mathrm{T}} : \boldsymbol{e}$ means $\sigma^0_{ij,k}e_{kj}$. Milton et al. (2006) also realized that if the microstructure in homogenization is sufficiently small, then these non-local effective operators in Eq. (1) might be approximated to local operators. In this sense, Eq. (9) can be understood as the limit result when the microstructure approaches to a point and the non-local effect of inhomogeneous media is represented by the gradient of pre-stresses.

The requirement of form-invariance in transformation methods is directly related to the Principle of Material Frame Indifference (PMFI). According to the detailed historic reviews (Truesdell and Noll, 2004; Frewer, 2009), the classical PMFI is related to the invariance of constitutive equations under transformations involving rigid movements and can be stated in two forms:

(a) The Hooke-Poisson-Cauchy form: constitutive equations must be invariant under a superimposed rigid rotation of the material.

(b) The Zeremba-Jaumann form: constitutive equations must be invariant under an

arbitrary change of the observer.

In form (a), the frame of reference is fixed and the state of the system is transformed, which is called the Active Transformation (AT). In form (b), the frame of reference is transformed while the state of the system is fixed, which is called the Passive Transformation (PT) or the *coordinate transformation*. Mathematically, every PT is equal to a corresponding AT but the opposite is not true (Frewer, 2009). Under a PT, the *form-invariance* can be naturally satisfied for tensor equations. The situation is more complex under an AT that cannot be converted to a PT, such as the aforementioned transformation used to design metamaterials. The *form-invariance* can be satisfied for active symmetry transformations that are related to certain conservation laws according to the Noether's theorems (Olver, 1993; Kienzler and Herrmann, 2000). Otherwise, new equations will be obtained after a general AT, for example Eq. (9), and must be verified by experiments before the implementation (Frewer, 2009). In addition, the *frame-independence* requires that transformed equations are independent of all properties of the transformation, which is a more restrictive requirement than the *form-invariance*. Since there is no equivalence principle of non-inertial reference frames, it is not reasonable to impose the *frame-independence* on a physical law in non-inertial systems.

Since the Willis equations could have very important potential applications for inhomogeneous media, such as composites, rocks, etc., it was worthwhile to design a corresponding verification experiment. However, there is no such experimental verification so far, because it is very difficult to quantitatively compare experimental measurements with the predictions from the Willis equations in general forms. For example, it is extremely difficult to accurately construct a three-dimensional or even a two-dimensional inhomogeneous pre-stress field in an experiment to verify Eq. (9). To circumvent this difficulty, this paper tries to check the validity of a simplified version of Eq. (9) in one-dimensional (1D) forms. For this purpose and inspired by the Hooke-Poisson-Cauchy form of PMFI, an experiment was conducted to investigate the behavior of a heavy soft spring in steady circular motion. The pretension in this rotational spring is inhomogeneous and can be accurately defined.

This paper is organized as follows. Section 2 firstly derives the 1D versions of Eq. (9) through the energy approaches. Then it rigorously proves that these 1D Willis-form equations can accurately describe the behavior of a rotational heavy spring, while in this case the traditional Hooke's law violates the PMFI. The theoretical predictions can be clearly supported by experimental results presented in Section 3. Detailed illustrations are also presented in this section to show that although these Willis-form equations are linear, they can naturally describe finite deformation problems, such as the spin-softening effect. Based on these findings, the final conclusions are given in Section 4.

## 2. Theoretical results

### *2.1. The 1D static Willis-form equations with displacement coupling terms*

The previous section has proved that the Willis-form equations with the displacement coupling terms are mathematically equivalent to the original Willis equations with the velocity coupling terms. Therefore, with the presence of pre-stresses and the ignorance of the inertia, one can use Eq. (9) to describe the static responses of inhomogeneous media. This part aims at giving a physical explanation to the 1D form of Eq. (9) through energy approaches.

Supposing there is a pre-stressed inhomogeneous bar in the initial configuration $B^0$. This bar elastically deforms to the configuration $B^1$ subjected to a small perturbation. During this process, a material point *X* on this bar moves from position $x^0$ in $B^0$ to position $x^1$ in $B^1$, with a small displacement $u = x^1 - x^0$ and a small strain $e = \mathrm{d}u / \mathrm{d}x^0$.

Since the bar is inhomogeneous, the strain energy density *W* varies in space and can be represented as $W = W(x, e)$ according to Kienzler and Herrmann (2000). Since everything should be expressed in the initial configuration $B^0$ in the linear elasticity theory, the strain energy density of *X* in $B^1$ can be represented approximately by using Taylor expansion as:

$$\begin{aligned} W\left(x^1, e\right) \approx W\left(x^0, 0\right) + \left.\frac{\partial W}{\partial x}\right|_{e=0,\, x=x^0} u + \left.\frac{\partial W}{\partial e}\right|_{e=0,\, x=x^0} e \\ + \frac{1}{2}\left.\frac{\partial^2 W}{\partial x^2}\right|_{e=0,\, x=x^0} u^2 + \frac{1}{2}\left.\frac{\partial^2 W}{\partial e^2}\right|_{e=0,\, x=x^0} e^2 + \left.\frac{\partial^2 W}{\partial x \partial e}\right|_{e=0, x=x^0} eu \end{aligned} . \tag{10}$$

In the initial configuration $B^0$, $W\left(x^0, 0\right)$, $\partial W / \partial x|_{e=0,\, x=x^0}$ and $\partial^2 W / \partial x^2|_{e=0,\, x=x^0}$ are customarily set to zeros, and $\partial W / \partial e|_{e=0,\, x=x^0}$ represents the pre-stress $\sigma^0\left(x^0\right)$. Therefore, Eq. (10) can be simplified as:

$$W\left(x^1, e\right) \approx \sigma^0\left(x^0\right) e + \frac{1}{2} C^0\left(x^0\right) e^2 + \frac{\mathrm{d}\sigma^0\left(x^0\right)}{\mathrm{d}x^0} eu , \tag{11}$$

where, $C^0\left(x^0\right) = \partial^2 W / \partial e^2|_{e=0, x=x^0}$ represents the tangent stiffness.

According to Eq. (11), the Cauchy stresses of *X* in $B^1$ can be calculated as

$$\sigma^1\left(x^1\right)=\frac{\partial W\left(x^1,e\right)}{\partial e}\approx\sigma^0\left(x^0\right)+C^0\left(x^0\right)e+\frac{\mathrm{d}\sigma^0\left(x^0\right)}{\mathrm{d}x^0}u\,. \tag{12}$$

Therefore, one can obtain the incremental Cauchy stress of $X$

$$\sigma=\sigma^1\left(x^1\right)-\sigma^0\left(x^0\right)\approx C^0\left(x^0\right)e+\frac{\mathrm{d}\sigma^0\left(x^0\right)}{\mathrm{d}x^0}u\,, \tag{13}$$

which is exactly the 1D version of Eq. (9a).

In configuration $B^0$, the equilibrium equation of $X$ can be written as:

$$\frac{\mathrm{d}\sigma^0\left(x^0\right)}{\mathrm{d}x^0}+f_b^0\left(x^0\right)=0\,, \tag{14}$$

where $f_b^0\left(x^0\right)$ denotes the body force and represents the inhomogeneity of $\sigma^0\left(x^0\right)$.

In configuration $B^1$, the equilibrium equation of $X$ can be written as:

$$\frac{\mathrm{d}\sigma^1\left(x^1\right)}{\mathrm{d}x^1}+f_b^1\left(x^1\right)=0\,, \tag{15}$$

where $f_b^1\left(x^1\right)$ denotes the body force and represents the inhomogeneity of $\sigma^1\left(x^1\right)$.

Since the perturbation is very small, one can assume the pre-stress does not change from $B^0$ to $B^1$. Therefore, the potential density of the body force due to the pre-stress and an additional body force $f^a$ can be written as:

$$V\left(x^1\right)=-\left[f_b^0\left(x^0\right)u+\frac{1}{2}\frac{\mathrm{d}f_b^0\left(x^0\right)}{\mathrm{d}x^0}u^2+f^a u\right]. \tag{16}$$

Then, one can construct the Lagrangian function $L=-W\left(x^1,e\right)-V\left(x^1\right)$ and obtains the incremental equation of motion, which is identical to the Euler-Lagrange equation (Kienzler and Herrmann, 2000):

$$\frac{\partial L}{\partial u}-\frac{\mathrm{d}}{\mathrm{d}x^0}\frac{\partial L}{\partial e}=0\,. \tag{17}$$

Substituting Eqs. (11) and (16) into Eq. (17), and noticing Eqs. (13) and (14), it obtains:

$$\begin{aligned}&\frac{\mathrm{d}\sigma\left(x^0\right)}{\mathrm{d}x^0}+\frac{\mathrm{d}\sigma^0\left(x^0\right)}{\mathrm{d}x^0}+f_b^0\left(x^0\right)+\frac{\mathrm{d}f_b^0\left(x^0\right)}{\mathrm{d}x^0}u+f^a\approx\frac{\mathrm{d}\sigma^0\left(x^0\right)}{\mathrm{d}x^0}e\\&\frac{\mathrm{d}\sigma\left(x^0\right)}{\mathrm{d}x^0}+f^a\approx\frac{\mathrm{d}\sigma^0\left(x^0\right)}{\mathrm{d}x^0}e+\frac{\mathrm{d}^2\sigma^0\left(x^0\right)}{\left(\mathrm{d}x^0\right)^2}u\end{aligned}\quad, \tag{18}$$

which is in accordance with Eq. (4b) without the inertia.

In the linear elasticity theory, $\mathrm{d}\sigma^1/\mathrm{d}x^1$ in Eq. (15) should be expressed in $B^0$:

$$\frac{\mathrm{d}\sigma^1}{\mathrm{d}x^1}=\frac{\mathrm{d}\left(x^1-u\right)}{\mathrm{d}x^1}\frac{\mathrm{d}\left(\sigma^0+\sigma\right)}{\mathrm{d}x^0}\approx\left(1-e\right)\left(\frac{\mathrm{d}\sigma^0}{\mathrm{d}x^0}+\frac{\mathrm{d}\sigma}{\mathrm{d}x^0}\right). \tag{19}$$

Since $e<<1$, it can be directly omitted from Eq. (19) if the effect of geometric nonlinearity during the change of configurations is ignored. However, if the geometric nonlinearity is taken into account, one should preserve $e$ in Eq. (19), which results in an effective body force $-\left(\mathrm{d}\sigma^0/\mathrm{d}x^0+\mathrm{d}\sigma/\mathrm{d}x^0\right)e\approx-\left(\mathrm{d}\sigma^0/\mathrm{d}x^0\right)e$ in Eq. (15) and consequently an additional potential density of body force $V^{nl}=\left(\mathrm{d}\sigma^0/\mathrm{d}x^0\right)eu$. Substituting $L^{nl}=-W\left(x^1,e\right)-V\left(x^1\right)-V^{nl}$ into Eq. (17), one obtains:

$$\frac{\mathrm{d}\sigma}{\mathrm{d}x^0}+f^a\approx 2\frac{\mathrm{d}\sigma^0}{\mathrm{d}x^0}e+\frac{\mathrm{d}^2\sigma^0\left(x^0\right)}{\left(\mathrm{d}x^0\right)^2}u\,. \tag{20}$$

The first right hand side term $\left(\boldsymbol{\sigma}^0\nabla\right)^{\mathrm{T}}:\boldsymbol{e}$ in Eq. (9b) is due to geometric nonlinearity in accordance with $\left(\mathrm{d}\sigma^0/\mathrm{d}x^0\right)e$ in Eq. (18). The second right hand side term $\boldsymbol{K}\cdot\boldsymbol{u}$ equals to $-\left(f_b^1-f_b^0\right)$ for this 1D problem. Since $f_b^1=\partial L/\partial u$, one can easily prove $-\left(f_b^1-f_b^0\right)=\left(\mathrm{d}\sigma^0/\mathrm{d}x^0\right)e+\left[\mathrm{d}^2\sigma^0/\left(\mathrm{d}x^0\right)^2\right]u$ according to Eqs. (11) and (16). Therefore, Eq. (20) is exactly the 1D version of Eq. (9b), which will be illustrated in more detail in Section 2.4.

### *2.2. The general formulation of a rotational spring*

As mentioned in Section 1, the pretension of a steadily rotating heavy soft spring is inhomogeneous and can be accurately defined. Therefore, it is possible to conduct a corresponding experiment to verify the 1D versions of Eq. (9), i.e., Eqs. (13) and (20). For this purpose, one can establish general formulations by using the theoretical model shown in Fig. 1.

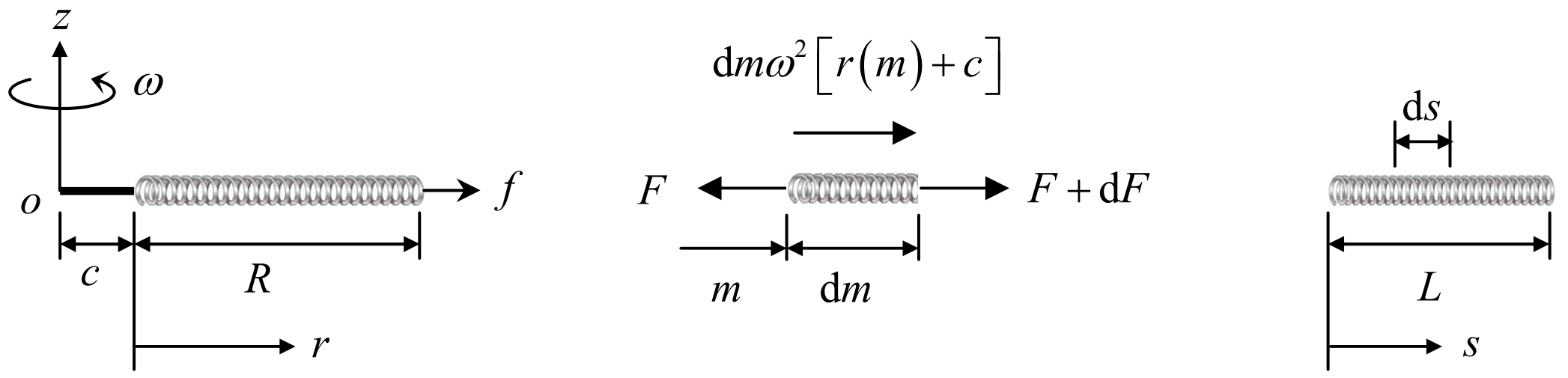

(a) rotational spring (b) a portion of rotational spring (c) unstretched spring

**Fig. 1.** The theoretical model.

As shown in Fig. 1 (a), one end of a rigid bar of length *c* connects with a uniform helical spring of total mass *M*, unstretched length *L* and spring constant *K*; another end of this rigid bar is fixed at the origin of a system that rotates at a constant angular velocity *ω*. In following deductions, only the static radial stretch of the spring under the end pull force *f* is interested in, regardless of the influences from friction, gravitational force and lateral bending.

Considering the equilibrium of a small portion of the spring in the deformed state (Fig. 1(b)), one can find the following relation:

$$\frac{\mathrm{d}F}{\mathrm{d}m} + \omega^2\left[r(m) + c\right] = 0, \tag{21}$$

where *m* ($0 \le m \le M$) is the material (Lagrangian) coordinate; $r(m)$ is the length from the inner end of the spring to this material point *m*; and *F* is the internal tensile force.

Denote the unstretched length of d$m$ portion of the spring as d$s$ (Fig. 1(c)), which satisfies $\mathrm{d}s/L=\mathrm{d}m/M$, because the mass of the unstretched spring is uniformly distributed. It is easy to prove that the corresponding spring constant of this uniform portion is $KL/\mathrm{d}s$. Since this portion deforms from the uniform unstretched state free of pretension, one can use the Hooke's Law to obtain:

$$F(m)=\frac{L}{\mathrm{d}s}K(\mathrm{d}r-\mathrm{d}s)=K\left(M\frac{\mathrm{d}r}{\mathrm{d}m}-L\right). \tag{22}$$

Substituting Eq. (22) into Eq. (21), yields:

$$\frac{\mathrm{d}^2 r}{\mathrm{d}m^2}+\frac{\omega^2}{MK}(r+c)=0. \tag{23}$$

Noticing the boundary conditions of $r(0)=0$ and $F(M)=f$, one can easily find the solution of Eq. (23):

$$r(m)=\left[\frac{f+KL}{Kq\cos(q)}+c\tan(q)\right]\sin\left(\frac{q}{M}m\right)+c\left[\cos\left(\frac{q}{M}m\right)-1\right], \tag{24a}$$

where

$$q=\omega\sqrt{\frac{M}{K}}. \tag{24b}$$

Eq. (24) explicitly shows that this rotational spring is inhomogeneous.

Substituting Eq. (24a) into Eq. (22), yields:

$$F(m)=\left[\frac{f+KL}{\cos(q)}+cKq\tan(q)\right]\cos\left(\frac{q}{M}m\right)-cKq\sin\left(\frac{q}{M}m\right)-KL. \tag{25}$$

One can notice that Eq. (21) is the equation of motion established on the deformed state, and Eq. (22) is the constitutive equation free of pretensions. Therefore, Eq. (23) is similar to the Lamé-Navier equation established on the deformed state, so that Eqs. (24) and (25) are the exact solutions of this finite deformation problem.

Assuming $q^2 << 1$, Weinstock (1964) used the similar model to explain why the equation of the period of a spring-mass system $2\pi\sqrt{M^a / K}$ should be corrected by using a modified end mass $M^{a*} = M^a + M / 3$, where $M^a$ is the real attached end mass. However, regardless of this assumption, one can exactly obtain the 1D Willis-form equations from Eqs. (24) and (25) as follows.

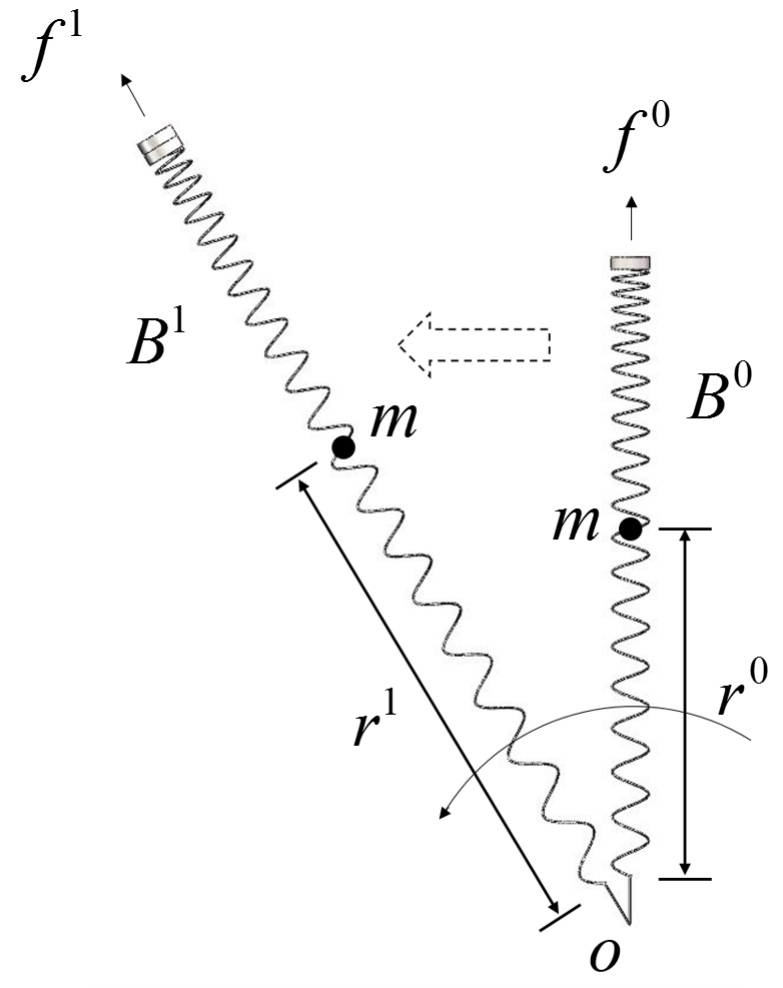

**Fig. 2.** The perturbation of the pre-stressed spring.

Since the Willis-form equations involve the incremental stress of a pre-stressed system, one should investigate the incremental tensile force of a rotational spring with pretension. For this purpose, one can focus on a material point *m* in the spring that rotates at constant angular velocity *ω*. As Fig. 2 shows, when the end pull force of this spring changes from $f^0$ to $f^1 = f^0 + \Delta f$ after a small perturbation of $\Delta f$, the position of this material point changes from the undeformed position $r^0$ (corresponding to the initial configuration $B^0$ in Section 2.1) to the deformed position $r^1 = r^0 + u$ (corresponding to the deformed configuration $B^1$ in Section 2.1) in the local coordinate system attached to the spring, and accordingly the internal tensile force at this material point changes from $F^0$ to $F^1 = F^0 + \Delta F$. The next two sections try to check if the relation between $\Delta F$ and $u$ and the equation of motion with respect to $\Delta F$ at the undeformed position $r^0$ are in Willis forms.

*2.3. The constitutive equation in Willis form*

According to Eqs. (24a) and (25), one can obtain the incremental tensile force at material point *m* as:

$$\Delta F = K_\omega u , \tag{26a}$$

where

$$K_\omega = q \cot\left(\frac{q}{M} m\right) K \tag{26b}$$

denotes the effective spring constant at angular velocity *ω*.

Eq. (26) is the exact solution of this finite deformation problem, which is defined at the deformed position $r^1$. However, $r^1$ is unknown in the linear approximation model, in which everything should be defined at the undeformed position $r^0$. For this purpose, the tensile force of material point *m* at $r^0$ can be represented as:

$$F^1\left(r^1\right) \approx F^1\left(r^0\right) + \frac{\mathrm{d}F^1\left(r^0\right)}{\mathrm{d}r^0} u , \tag{27}$$

where $F^1\left(r^0\right) = F^1\left(r^0, m', f^1\right)$ is the tensile force of an adjacent material point $m'$ at position $r^0$ in $B^1$. This means the tensile force of *m* is closely related to the tensile force of $m'$, which clearly demonstrates the non-local property of inhomogeneous media.

Since the existing pretension $F^0\left(r^0\right)$ at $r^0$ is independent of the small perturbation $\Delta f$, $F^1\left(r^0, m', f^1\right)$ equals to an incremental tensile force superimposed

on the pretension:

$$F^1\left(r^0, m', f^1\right) = F^0\left(r^0\right) + K_t u\,, \tag{28}$$

where $K_t$ represents the tangent spring constant at $r^0$ in $B^0$, and $K_t u$ is the incremental tensile force due to the change of the configuration induced by the small perturbation $\Delta f$:

$$K_t u = \left(\frac{\partial F}{\partial f} + \frac{\partial F}{\partial m}\frac{\partial m}{\partial f}\right)_{r=r^0, f=f^0} \Delta f\,, \tag{29}$$

where $\left(\partial F / \partial f\right)\Delta f$ represents the direct increment due to the change of end pull force *f*; and $\left(\partial F / \partial m\right)\left(\partial m / \partial f\right)\Delta f$ represents the indirect increment due to the change of material point from *m* to $m'$ at $r^0$.

According to Eq. (25), at the given material point *m*,

$$\frac{\partial F}{\partial f} = \frac{\cos\left(\dfrac{q}{M}m\right)}{\cos(q)}; \tag{30a}$$

and at the given configuration, i.e., for the given *f*,

$$\frac{\partial F}{\partial m} = -\left\{\left[\frac{f + KL}{Kq\cos(q)} + c\tan(q)\right]\sin\left(\frac{q}{M}m\right) + c\cos\left(\frac{q}{M}m\right)\right\}\frac{q^2}{M}K\,. \tag{30b}$$

According to Eq. (24a), at the given spatial position,

$$\frac{\partial f}{\partial m} = \left\{c - \left[\frac{f + KL}{Kq\cos(q)} + c\tan(q)\right]\cot\left(\frac{q}{M}m\right)\right\}\frac{q^2 K}{M}\cos(q); \tag{30c}$$

and for the displacement *u* of the given material point *m*,

$$\Delta f = \frac{Kq\cos(q)}{\sin\left(\dfrac{q}{M}m\right)}u\,. \tag{30d}$$

Substituting Eq. (30) into Eq. (29), yields:

$$K_t = \frac{\frac{f^0 + KL}{Kq\cos(q)} + c\tan(q)}{\frac{f^0 + KL}{Kq\cos(q)} + c\left[\tan(q) - \tan\left(\frac{q}{M}m\right)\right]} \frac{qK}{\sin\left(\frac{q}{M}m\right)\cos\left(\frac{q}{M}m\right)}. \tag{31}$$

Substituting Eq. (28) into Eq. (27), yields:

$$\Delta F = F^1\left(r^1\right) - F^0\left(r^0\right) = K_t u + \frac{\mathrm{d}\left[F^0\left(r^0\right) + K_t u\right]}{\mathrm{d}r^0}u \approx K_t u + \frac{\mathrm{d}F^0\left(r^0\right)}{\mathrm{d}r^0}u, \tag{32}$$

where the gradient of the tensile force in $B^0$ is:

$$\frac{\mathrm{d}F^0\left(r^0\right)}{\mathrm{d}r^0} = \left(\frac{\partial F}{\partial m}\frac{\mathrm{d}m}{\mathrm{d}r}\right)_{f=f^0}. \tag{33}$$

At the given configuration, i.e., for the given *f*, according to Eq. (24) one can define:

$$B(f) = \frac{\mathrm{d}m}{\mathrm{d}r} = \frac{1}{\left[\frac{f + KL}{Kq\cos(q)} + c\tan(q)\right]\cos\left(\frac{q}{M}m\right)\frac{q}{M} - c\sin\left(\frac{q}{M}m\right)\frac{q}{M}}. \tag{34a}$$

According to Eq. (24), Eq. (30b) can be further simplified as

$$\frac{\partial F}{\partial m} = -\omega^2\left(r + c\right). \tag{34b}$$

Substituting Eq. (34) into Eq. (33), obtains:

$$\frac{\mathrm{d}F^0\left(r^0\right)}{\mathrm{d}r^0} = -\omega^2 B\left(f^0\right)\left(r^0 + c\right). \tag{35}$$

Eq. (32) is the linear relation defined at the undeformed position $r^0$. However, one can easily prove that it is a very accurate approximation of the nonlinear solution of Eq. (26), i.e.,

$$K_\omega = K_t + \frac{\mathrm{d}F^0\left(r^0\right)}{\mathrm{d}r^0}. \tag{36}$$

Denoting the cross-sectional area of the spring as *A*, the effective incremental Cauchy stress as $\sigma = \Delta F / A$, and the effective pre-stress as $\sigma^0 = F^0 / A$, Eq. (32) can be rewritten as:

$$\sigma \approx \frac{K_t u}{A} + \frac{\mathrm{d}\sigma^0}{\mathrm{d}r^0} u \,. \tag{37}$$

In addition, the displacements of the material point *m* and its adjacent material point $m'$ are $u(m) = r^1(m) - r^0(m)$ and $u'(m') = r'^1(m') - r'^0(m')$, respectively. Therefore, according to Eqs. (24a) and (34a), one can find the relation:

$$e = \frac{\mathrm{d}u}{\mathrm{d}r^0} = \lim_{m' \to m} \frac{u(m) - u'(m')}{r^0(m) - r'^0(m')} = B\left(f^0\right)\frac{q}{M}\cot\left(\frac{q}{M}m\right)u \,. \tag{38}$$

Substituting Eqs. (31), (34a) and (38) into Eq. (37), yields:

$$\sigma = C^0 e + \frac{\mathrm{d}\sigma^0}{\mathrm{d}r^0} u \,, \tag{39a}$$

where

$$C^0 = \frac{f^0 + KL + cKq\sin(q)}{A\cos(q)\cos\left(\frac{q}{M}m\right)} \,. \tag{39b}$$

This is exactly a 1D constitutive equation in Willis form, in accordance with Eq. (13).

Similar to Section 2.1, one can also obtain Eq. (39a) from the energy point of view, if noticing the strain energy density of material point *m* equals to the work done by the pre-stress $\sigma^0\left(r^1\right)$ and the deformation stress $C^0 e$ :

$$W = \sigma^0\left(r^1\right)e + \frac{1}{2}C^0 e^2 = \left[\sigma^0\left(r^0\right) + \frac{\mathrm{d}\sigma^0\left(r^0\right)}{\mathrm{d}r^0}u\right]e + \frac{1}{2}C^0 e^2 \,. \tag{40}$$

*2.4. The equation of motion in Willis form*

According to Eq. (35) and noticing $\sigma^0 = F^0 / A$ and $\sigma^1 = F^1 / A$, one can obtain the equations of motion of material point *m* in $B^0$ and $B^1$:

$$\frac{\mathrm{d}\sigma^0}{\mathrm{d}r^0} + \frac{\omega^2}{A} B\left(f^0\right)\left(r^0 + c\right) = 0 \,, \tag{41}$$

$$\frac{\mathrm{d}\sigma^1}{\mathrm{d}r^1}+\frac{\omega^2}{A}B\left(f^1\right)\left(r^1+c\right)=0. \tag{42}$$

In Eq. (42), $\frac{\mathrm{d}\sigma^1}{\mathrm{d}r^1}=\frac{\mathrm{d}r^0}{\mathrm{d}r^1}\frac{\mathrm{d}\sigma^1}{\mathrm{d}r^0}$, $\sigma^1=\sigma^0+\sigma$, $r^1=r^0+u$ and $B\left(f^1\right)$ can be denoted as $B\left(f^1\right)=B\left(f^0\right)+\Delta B$. If $\Delta f$ is very small, $\sigma$, $u$ and $\Delta B$ are also very small. Therefore, substituting these relations into Eq. (42) and noticing Eq. (41), one obtains

$$\begin{gathered}\left(1-\frac{\mathrm{d}u}{\mathrm{d}r^0}\right)\left(\frac{\mathrm{d}\sigma^0}{\mathrm{d}r^0}+\frac{\mathrm{d}\sigma}{\mathrm{d}r^0}\right)+\frac{\omega^2}{A}\left[B\left(f^0\right)+\Delta B\right]\left(r^0+u+c\right)=0 \\ \frac{\mathrm{d}\sigma}{\mathrm{d}r^0}\approx\frac{\mathrm{d}\sigma^0}{\mathrm{d}r^0}e-\frac{\omega^2}{A}\left[B\left(f^0\right)u+\left(r^0+c\right)\Delta B\right]\end{gathered}, \tag{43}$$

which is exactly the 1D version of Eq. (9b) free of the additional external body force. And

$$\Delta B\approx\left(\frac{\partial B}{\partial r^0}u+\frac{\partial B}{\partial f}\Delta f\right)_{f=f^0}, \tag{44}$$

where $\left(\partial B/\partial r^0\right)u$ represents the increment due to the change of position and $\left(\partial B/\partial f\right)\Delta f$ represents the increment due to the change of end pull force $f$.

According to Eq. (41), one can obtain:

$$\frac{\mathrm{d}^2\sigma^0}{\left(\mathrm{d}r^0\right)^2}u=-\frac{\omega^2}{A}\frac{\mathrm{d}\left[B\left(f^0\right)\left(r^0+c\right)\right]}{\mathrm{d}r^0}u=-\frac{\omega^2}{A}\left[B\left(f^0\right)+\left(r^0+c\right)\left.\frac{\partial B}{\partial r^0}\right|_{f=f^0}\right]u. \tag{45}$$

According to Eqs. (30d) and (34a), and referring to Eqs. (38) and (41), it is easy to prove:

$$\frac{\mathrm{d}\sigma^0}{\mathrm{d}r^0}e=\frac{\omega^2\left(r^0+c\right)}{A}\left.\frac{\partial B}{\partial f}\right|_{f=f^0}\Delta f. \tag{46}$$

Then, adding Eq. (45) with Eq. (46) and noticing Eq. (44), yields

$$\frac{\mathrm{d}^2\sigma^0}{\left(\mathrm{d}r^0\right)^2}u+\frac{\mathrm{d}\sigma^0}{\mathrm{d}r^0}e\approx-\frac{\omega^2}{A}\left[B\left(f^0\right)u+\left(r^0+c\right)\Delta B\right]. \tag{47}$$

Therefore, Eq. (43) can be written as:

$$\frac{\mathrm{d}\sigma}{\mathrm{d}r^0} \approx 2\frac{\mathrm{d}\sigma^0}{\mathrm{d}r^0}e + \frac{\mathrm{d}^2\sigma^0}{\left(\mathrm{d}r^0\right)^2}u, \tag{48}$$

which is in accordance with Eq. (20).

Similar to Section 2.1, Eq. (48) can also be obtained from the Euler-Lagrange equation, if noticing Eq. (40), $V = \left(\mathrm{d}\sigma^0/\mathrm{d}r^0\right)u + \left[\mathrm{d}^2\sigma^0/\left(\mathrm{d}r^0\right)^2\right]u^2/2$ and $V^{nl} = \left(\mathrm{d}\sigma^0/\mathrm{d}r^0\right)eu$.

*2.5. Discussions*

From Eq. (24a), one can obtain:

$$f = \left[q\cot(q) - 1\right]KL - c\left[\frac{\sin^2(q)}{\cos(q)} + \cos(q) - 1\right]q\cot(q)K + q\cot(q)K(R - L), \tag{49}$$

where $R = r(M)$ is the full length of the stretched spring. This equation has the different form as the Hooke's law $f = K(R-L)$, unless $q = 0$. In addition, based on Eq. (49) one can obtain the incremental relation $\Delta f = q\cot(q)K\Delta R$. Although it has the same form as the Hooke's law $\Delta f = K\Delta R$, the effective spring constant $q\cot(q)K$ contains *ω*, a property of the rotating frame, so that it is not frame independent (Frewer, 2009). All in all, this demonstrates that the Hooke's law violates the PMFI under a rotation transformation if the spring mass can not be ignored.

Eq. (24) reveals that the spring is generally inhomogeneous, so that it naturally follows the 1D Willis-form equations of (39a) and (48). These equations will degenerate to the Hooke's law if the rotation velocity is zero or the spring mass is ignored, because in these cases $q = 0$, and hence $K_t = K$ and $\mathrm{d}F^0/\mathrm{d}r^0 = 0$.

In practice, the end pull force $f$ can be conveniently generated by attaching an end mass $M^a$ at the outer end of the spring. In this way,

$$f = M^{a}\omega^{2}\left(c + R + d\right), \tag{50}$$

where $d$ is the distance between the outer end of the spring and the barycenter of the attached mass. The $\Delta f$ can be generated by either change the attached end mass $M$ or change the angular velocity $\omega$. Since $f$ is the function of the unknown $R$, this is a finite deformation problem, i.e., the $\Delta F$ in Eq. (26) is not only related to the known $r^0$ but also related to the unknown $u$ due to $\Delta F$. Usually, it is very difficult to analytically solve a finite deformation problem. Instead, one has to piecewisely linearize this nonlinear problem and uses only tangent predictions (e.g., calculate $\Delta f$ by Eq. (50) while using $R = R^0$) to obtain approximate solutions iteratively, which are always suffered from the well-known drifting effect. However, for this simple case of a rotational heavy spring, it does possible to obtain the analytical solutions. The $K_\omega$ in Eq. (26) can be regarded as the exact secant spring constant, which can be decomposed into the tangent spring constant $K_t$ plus the gradient of the tensile force $\mathrm{d}F^0/\mathrm{d}r^0$ according to Eq. (36). Consequently, Eq. (39a) is the 1D constitutive equation in Willis form. Experimental results in Section 3.2 will give further demonstrations of the importance of $\mathrm{d}F^0/\mathrm{d}r^0$, which can naturally explain the spin-softening effect. In addition, the linear equation of motion should be established on the undeformed state. Following this concept, one can obtain Eq. (48), the 1D equation of motion in Willis form. It is also a much accurate linear approximation of the equation of motion for the incremental stress on a pre-stressed medium. All in all, these Willis-form equations are very accurate linear approximations of the nonlinear equations for finite deformation problems.

Since the governing equations of a heavy spring change from the Hooke's Law to the Willis-form equations after a rotation transformation, it is necessary to verify these new equations with experimental results before the implementation (Frewer, 2009). For this purpose, experiments are conducted in Section 3 to verify Eqs. (26) and (49), because the Willis-form equations of (39a) and (48) are natural results from these two equations.

## 3. Experimental verifications

### *3.1. The experiment setup*

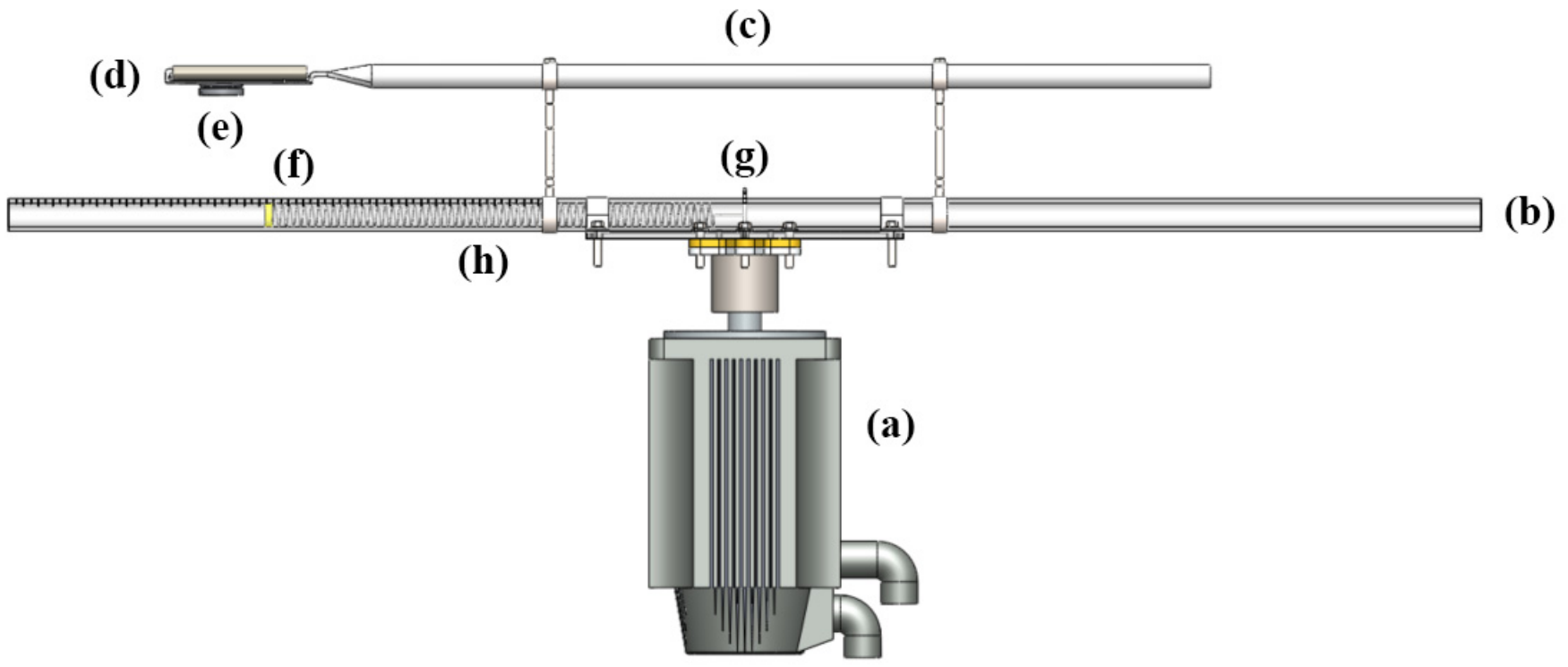

**Fig. 3.** (a) Servo motor. (b) Plexiglass tube with a Teflon film lined on the inner surface. (c) Selfie stick. (d) Smart phone. (e) Wide-angle lens. (f) Film ruler pasted on the outer surface of the plexiglass tube. (g) Dowel pin. (h) Spring-mass assemblage.

The experiment setup is illustrated in Fig. 3 and Fig. 4. A servo motor accurately drives a plexiglass tube to rotate at a specified constant angular velocity. A spring specimen with an attached end mass is put into the tube with one end fixed at the rotation center by a dowel pin. In this way, the gravitational force and aerodynamic drag can be ignored so that the spring can deform only in length direction without lateral bending. In addition, since the spring-mass assemblage is in a static state relative to the tube, the Coriolis effect is negligible. To reduce the friction force between the spring and the inner surface of the tube, the spring diameter (12mm) is set smaller than the inner diameter (15mm) of the tube to reduce the contact area, and the inner surface of the tube is lined with a thin Teflon film with a very small friction coefficient.

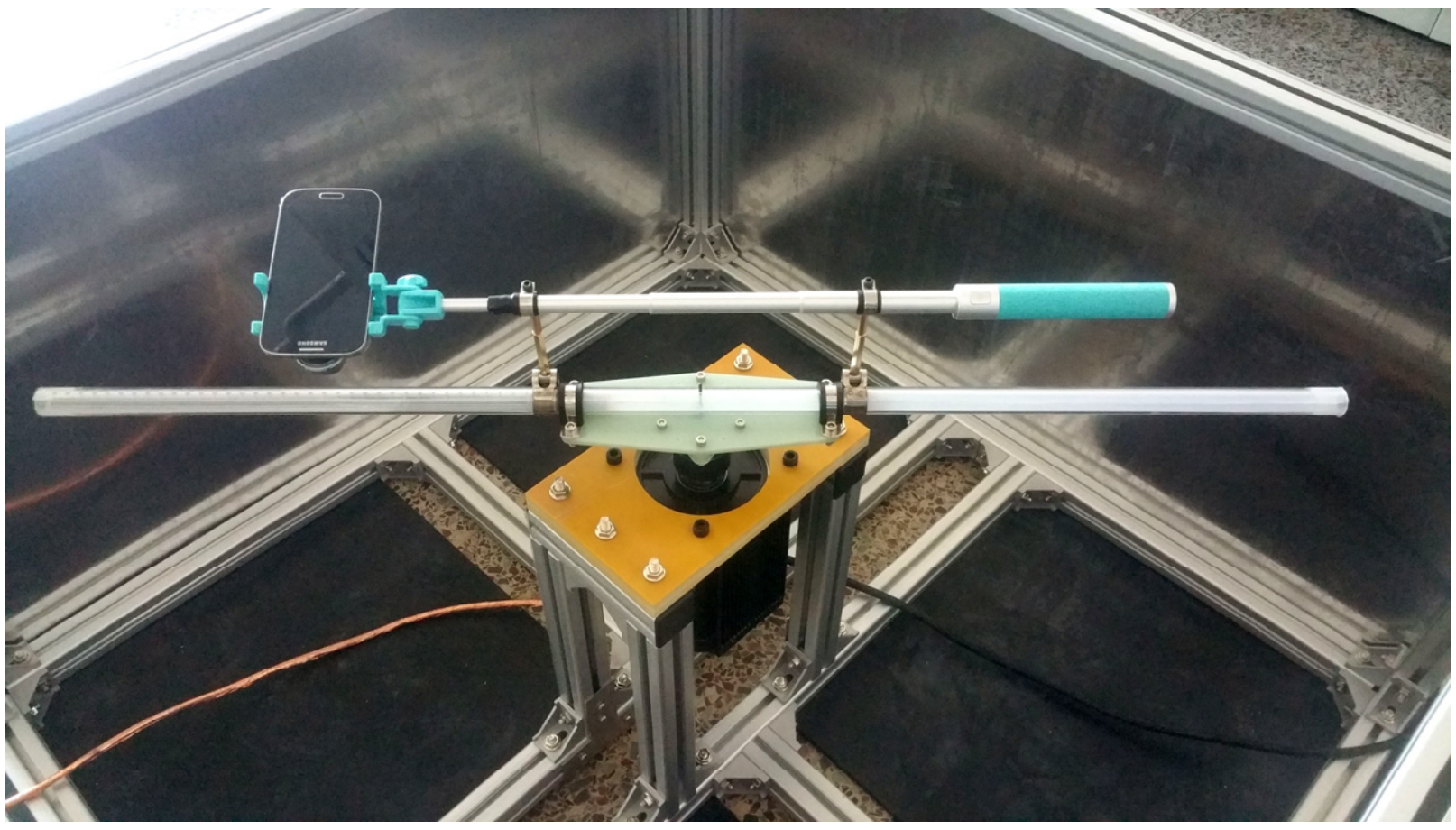

**Fig. 4.** The photo of the experiment setup.

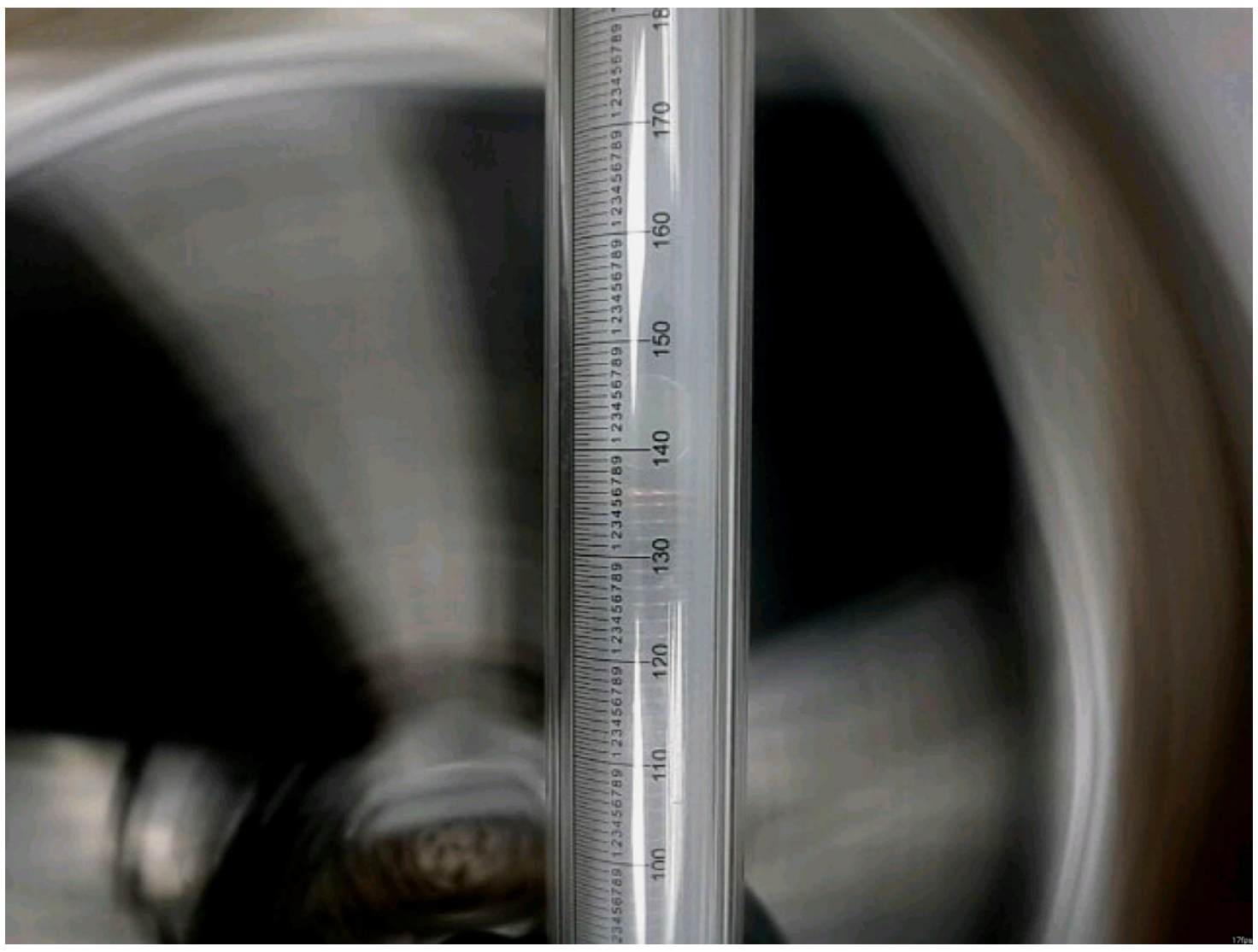

**Fig. 5.** A real-time image of the rotating spring-mass assemblage and the tube.

The image of the spring-mass assemblage and the tube is recorded in real time by a smart phone with a wide-angle lens, which is mounted on a selfie stick that rotates along with the tube. This image is wirelessly transmitted to another smart phone a distance away from the experiment setup through Wi-Fi. In this way, an observer can easily measure the spring length through a film ruler pasted on the outer surface of the tube (Fig. 5).

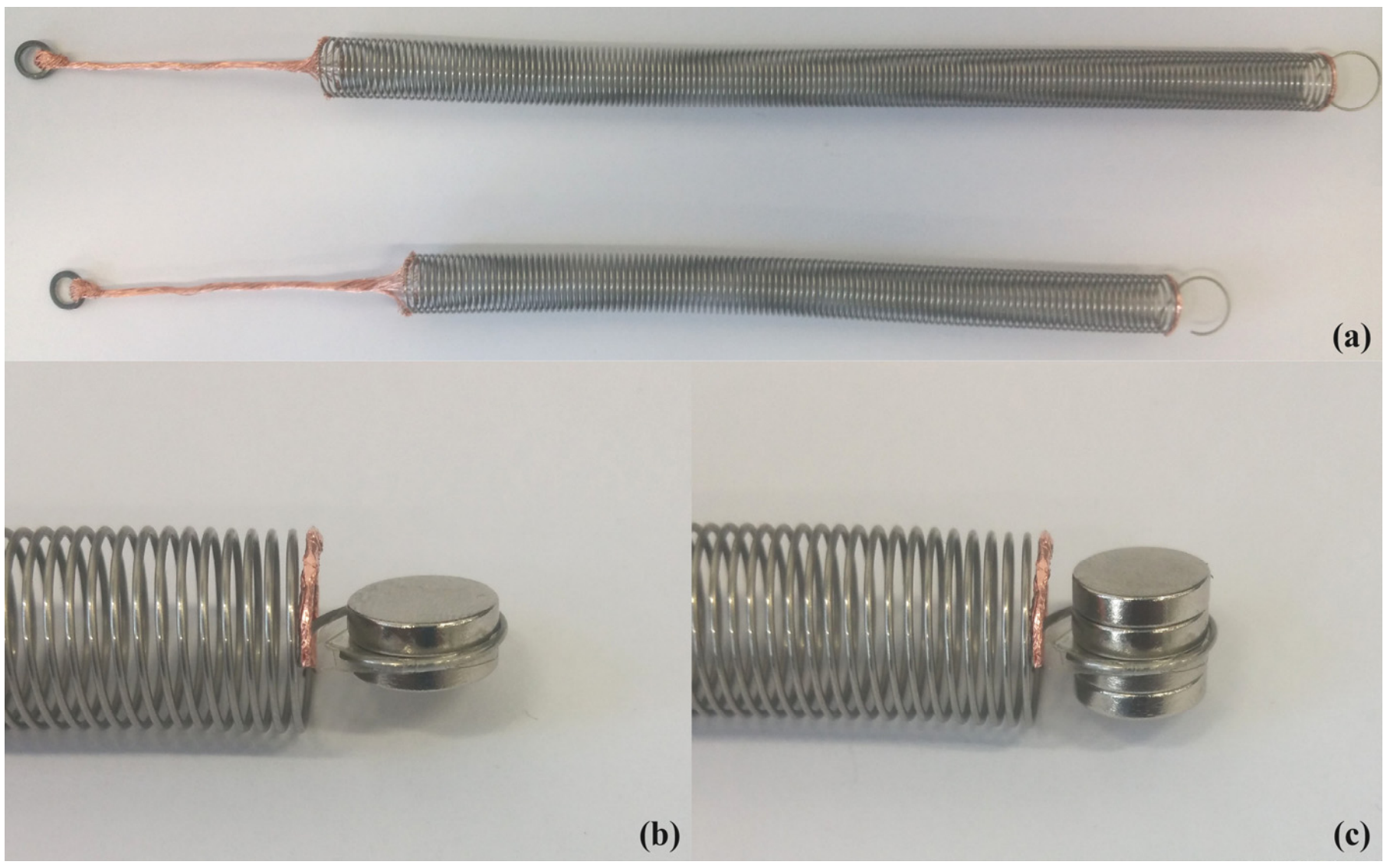

**Fig. 6.** (a) Two stainless steel springs with connection copper wires. (b) The attached end mass of $M^a = 2.37\text{g}$. (c) The attached end mass of $M^a = 4.75\text{g}$.

As Fig. 6 shows, two stainless steel springs are tested in the experiment. The diameter of spring wire and the ring gap are both approximately 0.6mm. The mass $M$ and the unstretched length $L$ of the two springs are 12.83g and 9.50g, 201.0mm and 153.0mm, respectively. A copper wire of length $c$ (Fig. 1) connects the inner end of the spring and a small ring, which is fixed at the rotation center by the dowel pin (Fig. 3). Some magnetic discs of mass $M^a$ are attached at the outer end of the spring, which generate the end pull force according to Eq. (50), where $d$ is 8.0mm and 7.0mm, and $c$ is 64.5mm and 72.0mm for the 201mm-length spring and the 153mm-length spring, respectively. Two attached end masses of $M^a = 2.37\text{g}$ and $M^a = 4.75\text{g}$ are used to generate end pull force $f^0$ and $f^1$ mentioned in Section 2, respectively. The upper half ring of the outer end of the spring is wrapped with a piece of copper foil, which serves as a highly reflective marker to facilitate reading the scale value (Fig. 5).

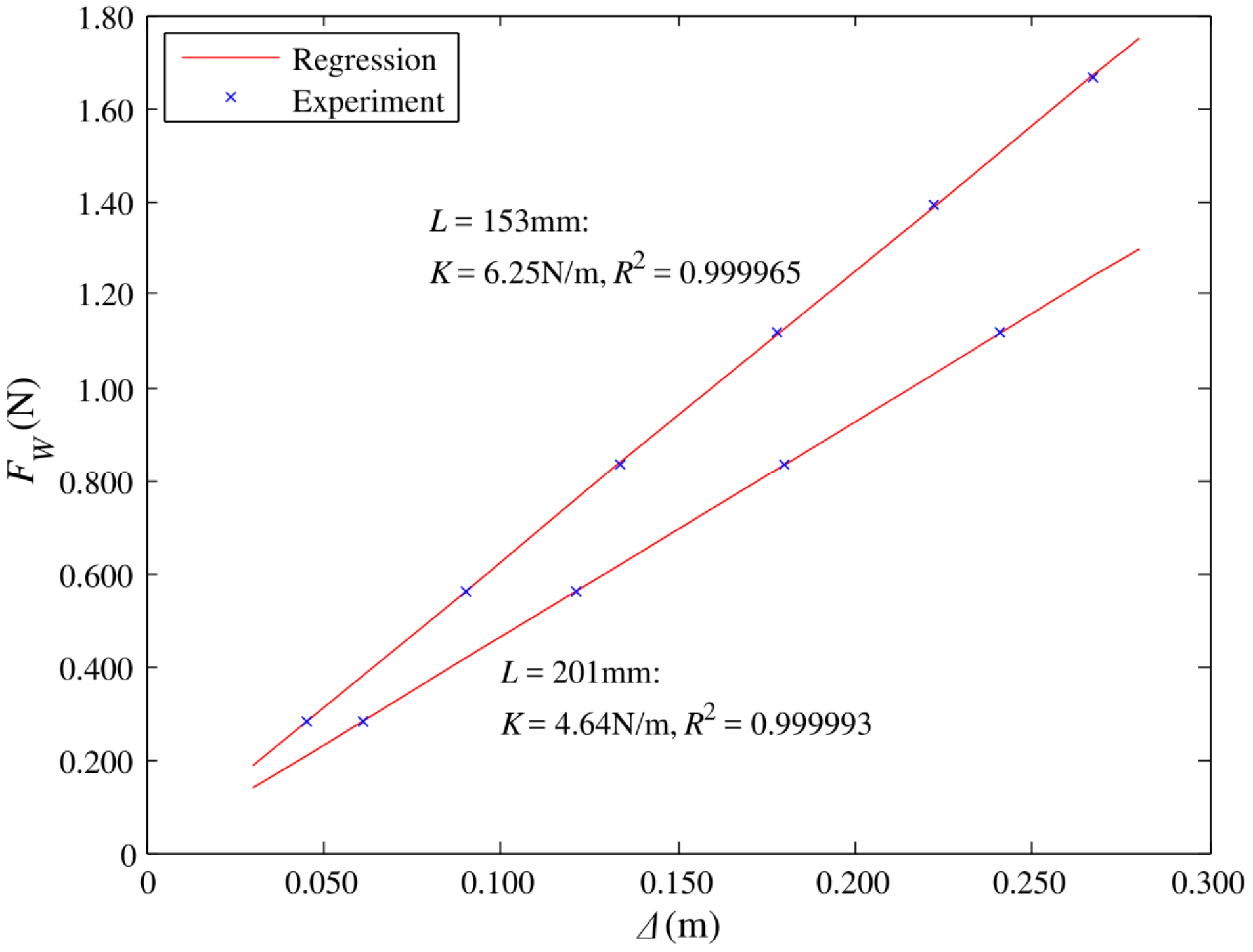

**Fig. 7.** The $F_W$ -$\Delta$ relations of the two springs.

Before the experiment, the spring constants were tested by measuring the deformation $\Delta$ of the hanging spring under the attached end weight force $F_W$. As Fig. 7 shows, the $F_W$-$\Delta$ relations of the two springs are sufficiently linear with very high coefficients of determination $R^2$. The finally obtained spring constants $K$ for the 201mm-length spring and the 153mm-length spring are 4.64 N/m and 6.25 N/m, respectively.

Each spring specimen in Fig. 6 is tested through the following steps:

*Step 1*: Place the spring-mass assemblage with the attached end mass of $M^a = 2.37\text{g}$ into the tube and fix it at the rotation center by the dowel pin;

*Step 2*: Turn on the servo motor and gradually speed it up to a relatively large angular velocity. Keep the constant angular velocity for a moment and then gradually slow down and turn off the servo motor. In this way, the spring is in a naturally unstretched state. Then, one can measure $c$ and $L$;

*Step 3*: Run the servo motor at a series of constant angular velocities and record the corresponding spring length from the real-time image in Fig. 5. The range of angular velocity is chosen such that the spring is sufficiently stretched within the length limit of the tube (0.5m in this experiment setup). To ensure the accuracy of the measurement, one should adjust the smart phone at a proper position so that the reflective marker locates in the optical non-distorted zone, which is about 10mm-width in the view center of the camera (Fig. 5). The resolution of the recorded spring length is estimated to 0.1mm;

*Step 4*: Repeat Step 1 through Step 3 for the spring-mass assemblage with the attached end mass of $M^a = 4.75\text{g}$.

*3.2. The experimental results*

Substituting Eq. (50) into Eq. (49), obtains the theoretical prediction of spring length:

$$R^{Theory}=\frac{p\sin(q)}{q\left[p\cos(q)-q\sin(q)\right]}L+\left[\frac{p}{p\cos(q)-q\sin(q)}-1\right]c+\frac{q\sin(q)}{p\cos(q)-q\sin(q)}d$$

, (51)

where $p=M/M^{a}$. This prediction is compared with experimental measurement $R^{Exp}$ in Fig. 8, which clearly shows that $R^{Exp}$ agree with $R^{Theory}$ extremely well for the two spring specimens. When the attached end mass is 4.75g, the maximum deviations ($\max\Delta_R$) of these two springs are both only 0.83% by chance.

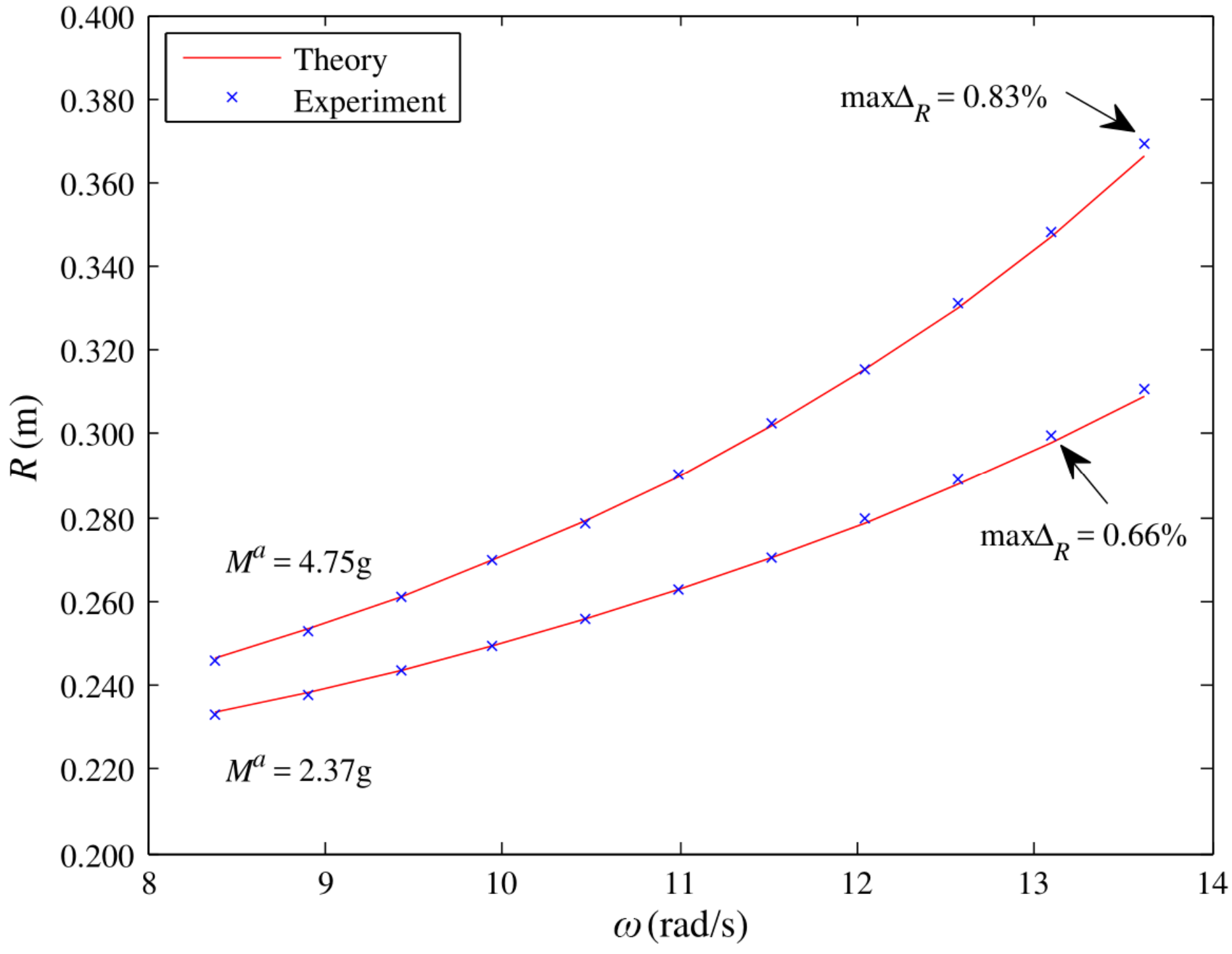

(a) the 201mm-length spring

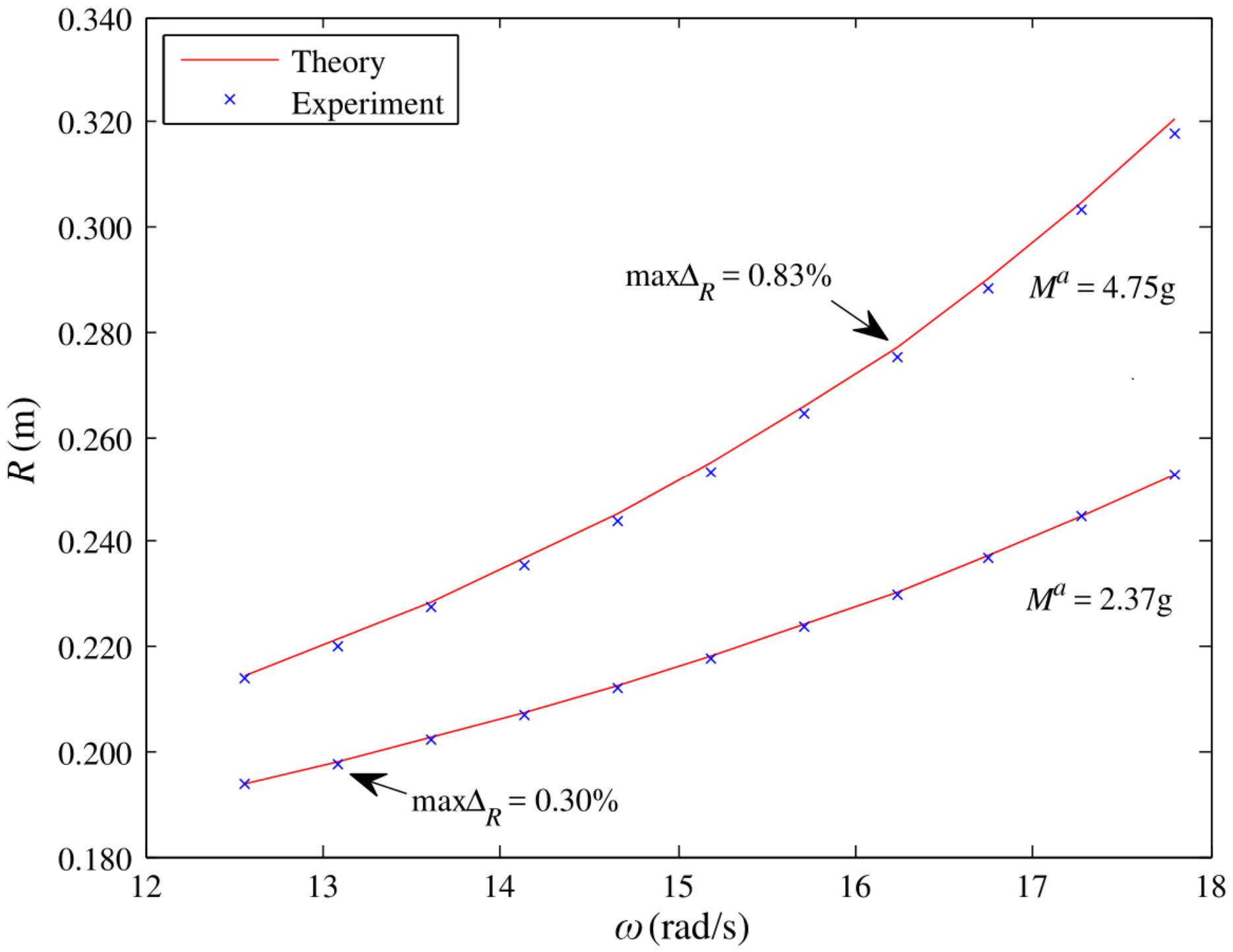

(b) the 153mm-length spring

**Fig. 8.** Comparisons between $R^{Exp}$ and $R^{Theory}$.

According to Eq. (26b), the theoretical effective spring constant at the outer end of the spring is $K_{\omega}^{Theory} = q\cot(q)K$. This value can be experimentally obtained by $K_{\omega}^{Exp} = (f^1 - f^0)/(R^1 - R^0)$ according to Eq. (50). The comparison between $K_{\omega}^{Exp}$ and $K_{\omega}^{Theory}$ in Fig. 9 shows larger deviations than those in Fig. 8. This is probably due to the limited resolution (0.1mm) of the denominator $R^1 - R^0$, which amplifies measurement errors. Since these two springs are made from the same wire with the same diameter and ring gap, their only difference is the total length *L*. Because the total friction force is proportional to *L* and the total centrifugal force contains the contribution of the added end mass, which is proportional to *R*. Since *R* > *L*, the friction force has more impact on the shorter spring. This can be confirmed by the curves in Fig. 8, where $R^{Theory}$ is always larger than $R^{Exp}$ for the 153mm-length spring. This is why $K_\omega^{Exp}$ is globally larger than $K_\omega^{Theory}$ for the 153mm-length spring in Fig. 9. In addition, it observes that the deviation of the 201mm-length spring is smaller than that of the 153mm-length spring. This is not only due to the aforementioned friction impact difference, but also because the longer spring is softer than the shorter spring, so that its $R^1 - R^0$ is relatively more accurate than that of the shorter spring. However, since the largest deviation is smaller than 4%, the experiment results are convictive enough to verify Eq. (26).

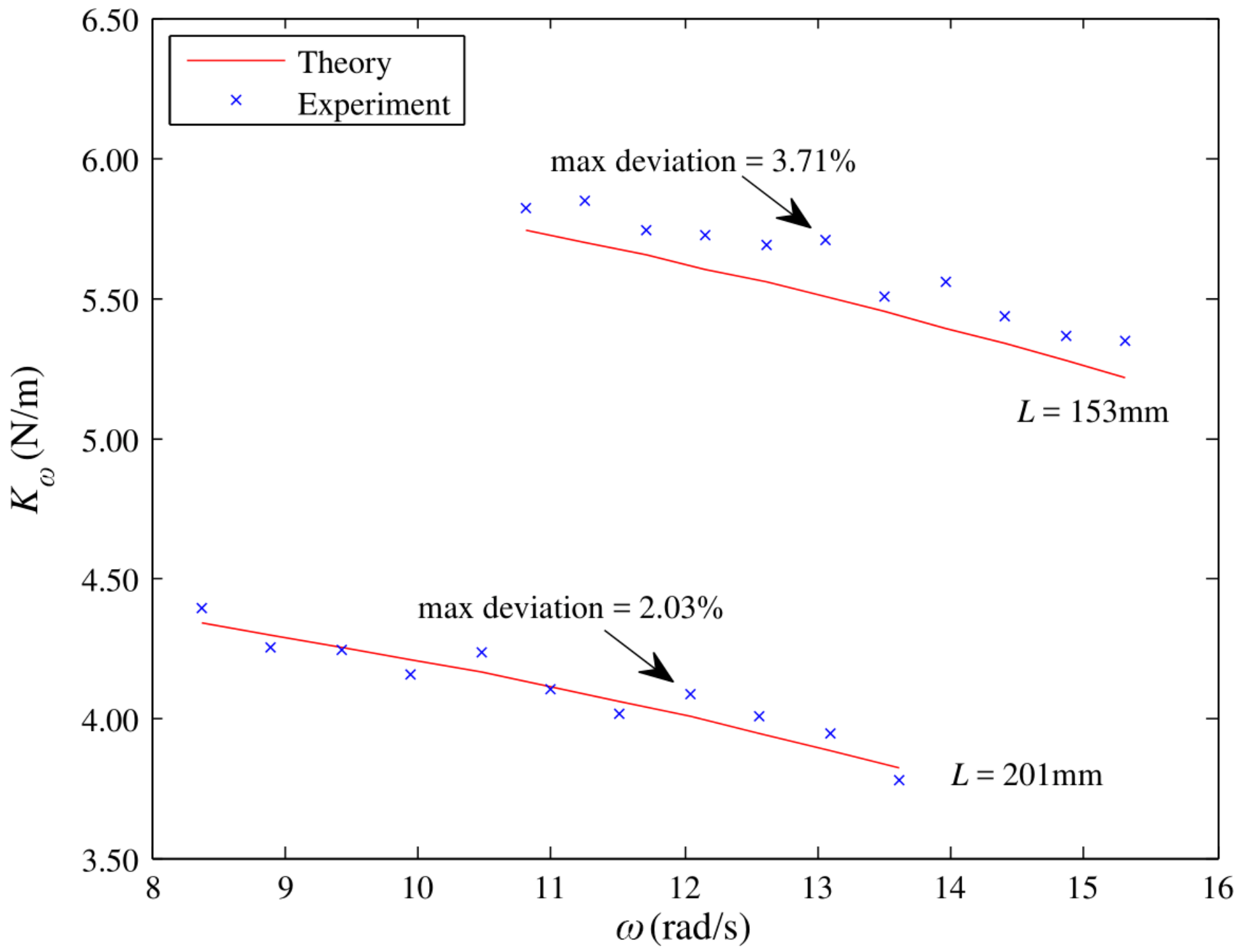

**Fig. 9.** Comparisons between $K_\omega^{Theory}$ and $K_\omega^{Exp}$.

To investigate the contribution of the additional stiffness due to the gradient of the tensile force in $K_\omega = K_t + \mathrm{d}F^0/\mathrm{d}r^0$, one can depict the distributions of the index $Q = -\left(\mathrm{d}F^0/\mathrm{d}r^0\right)/K_t$ at different rotation velocities while fixing $M^a = 2.37\mathrm{g}$ in Fig. 10 and Fig. 11. It observes that $Q$ gradually grows along the spring length at a given $\omega$;

$Q$ also increases as the rotation speeds up, and can reach a large value of 0.5 at the outer end of the 201mm-length spring; and $Q$ is larger for softer springs at the outer end.

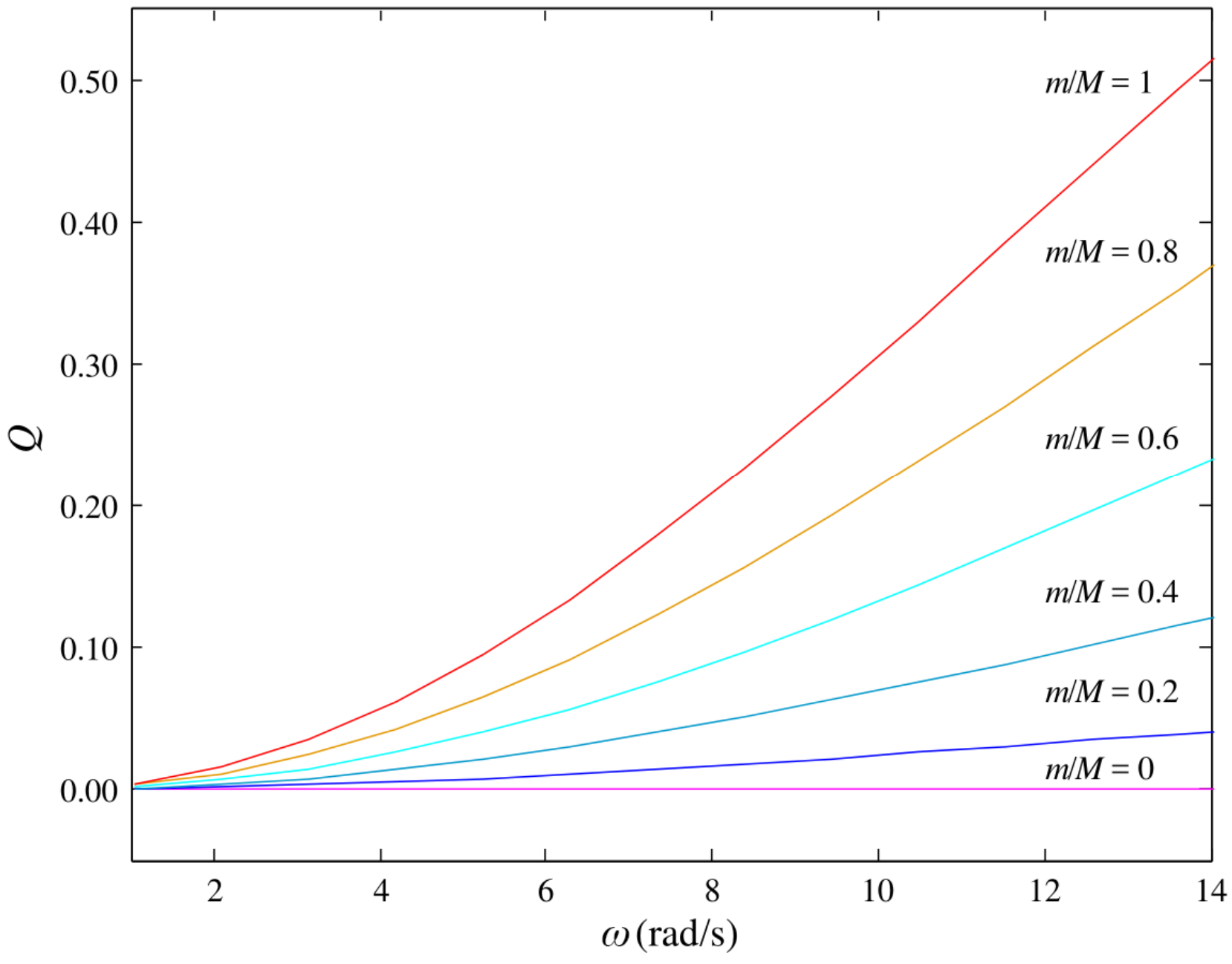

(a) the 201mm-length spring

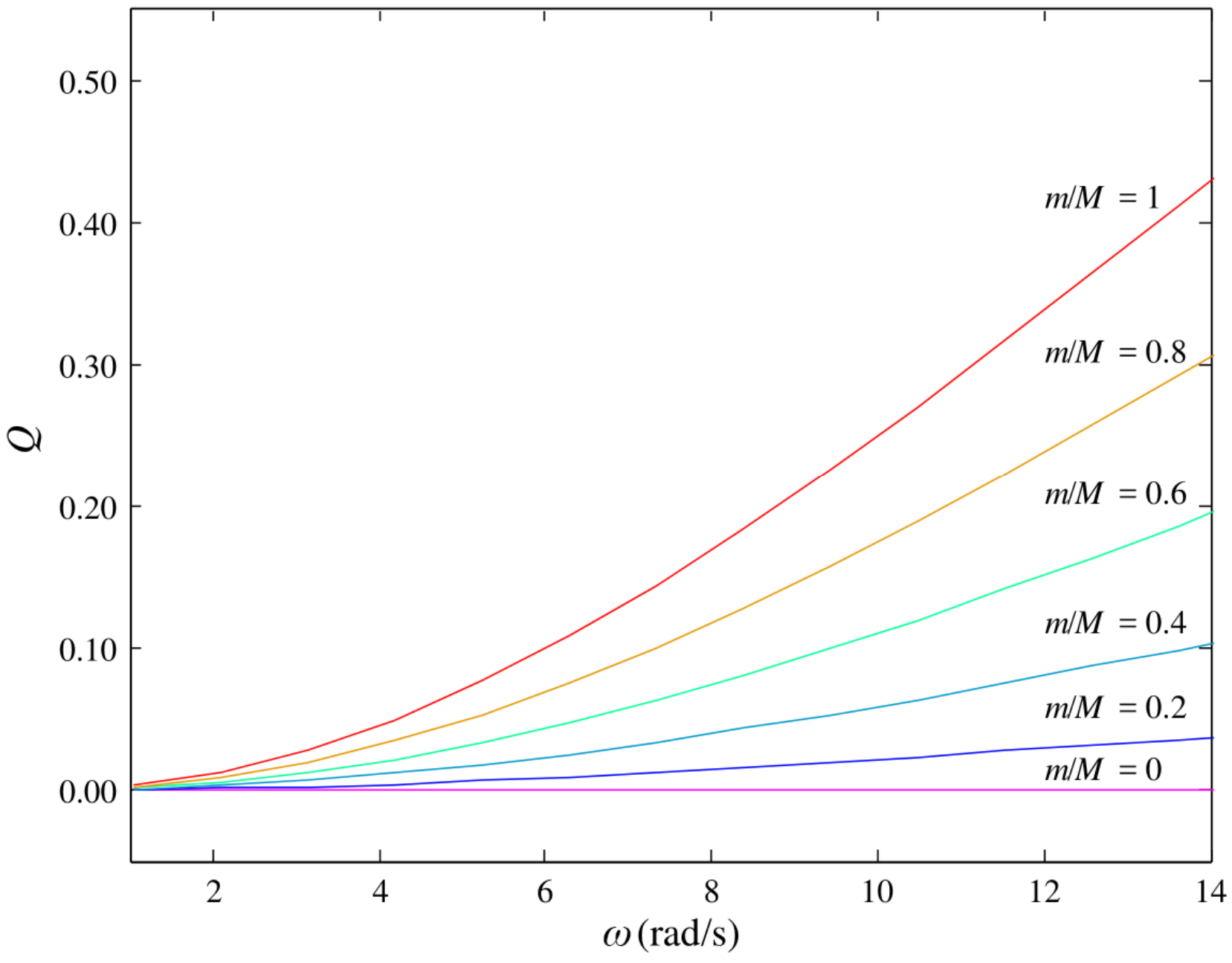

(b) the 153mm-length spring

**Fig. 10.** The distributions of $Q$ at different rotation velocities.

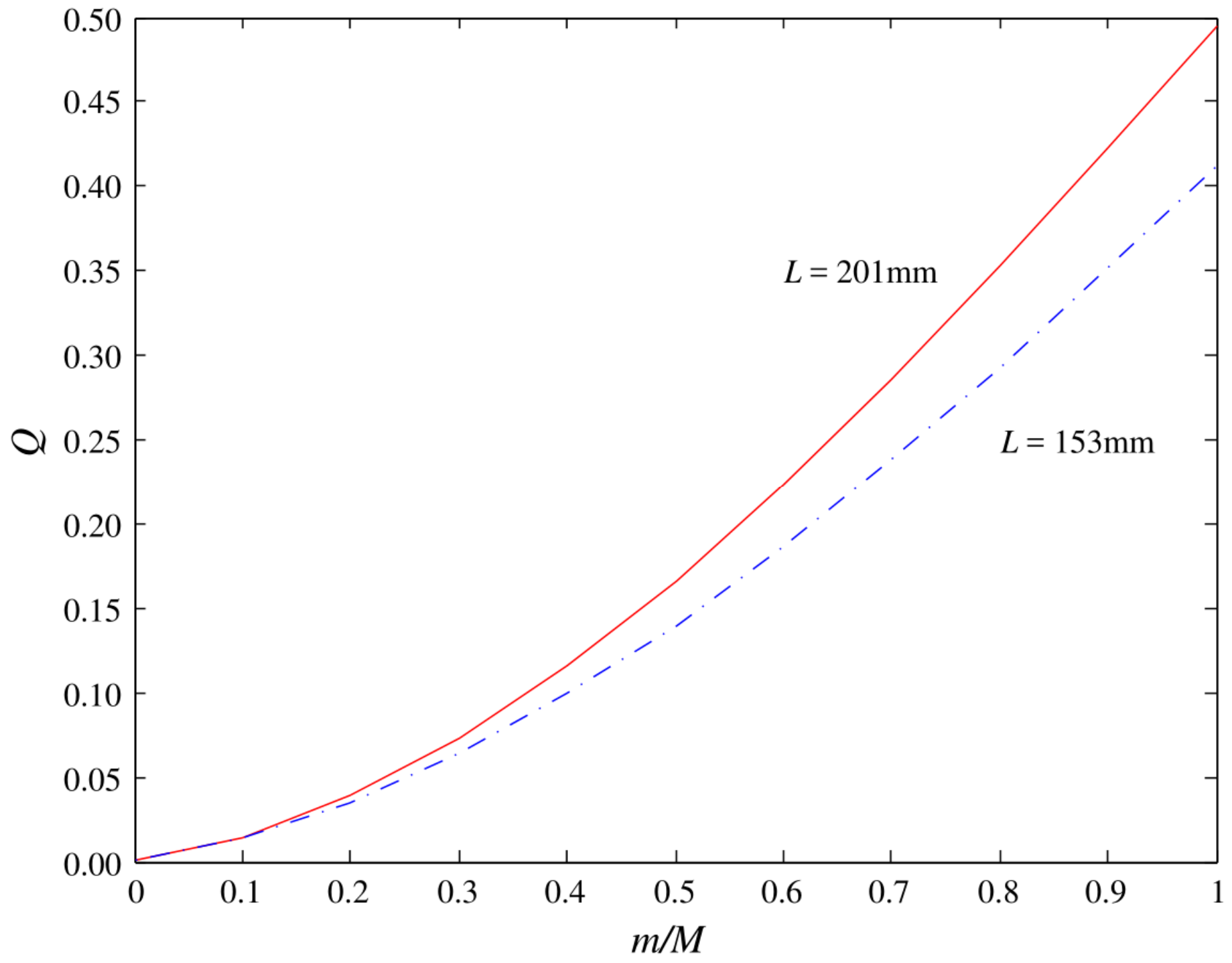

**Fig. 11.** The comparison of the distribution of $Q$ between the two springs at $\omega = 13.61\,\text{rad/s}$.

One can also notice in Fig. 9 that $K_\omega$ decreases when the rotation speeds up. This is a synthetic result from the stress-stiffening effect due to the increase of the centrifugal tensile force (Mayo et al., 2004) and the spin-softening effect, which represents the increase of centrifugal load with radial deformation (Mazière et al., 2009). To illustrate the contributions of these two effects to the final effective spring constant, $K_\omega$, $K_t$ and $\mathrm{d}F^0/\mathrm{d}r^0$ are compared in Fig. 12 and Fig. 13 when taking the same $M^a = 2.37\,\text{g}$. As Fig. 12 shows, since $K_t$ and $\mathrm{d}F^0/\mathrm{d}r^0$ decreases along the spring length at a given angular velocity, $K_\omega$ decreases more rapidly than $K_t$. As Fig. 13 shows, with the increase of $\omega$, the tangent spring constant $K_t$ increases due to the increase of the tensile force. This is a typical stress-stiffening effect. The spin-softening effect is represented by $\mathrm{d}F^0/\mathrm{d}r^0$, which decreases with the increase of $\omega$. Because the decrease of $\mathrm{d}F^0/\mathrm{d}r^0$ dominates the increase of $K_t$, the synthetic result $K_\omega$ shows the spin-softening effect. These observations demonstrate that $\mathrm{d}F^0/\mathrm{d}r^0$ plays an important role in the Willis-form equations, which can naturally describe the spin-softening effect. This also implies that although the Willis-form equations are linear, they can accurately predict geometrical nonlinear behaviors owing to the gradient of pre-stresses. Because of these features, the Willis-form equations can give a clear explanation why the rotational spring presents the spin-softening effect with the increase of internal tensile force.

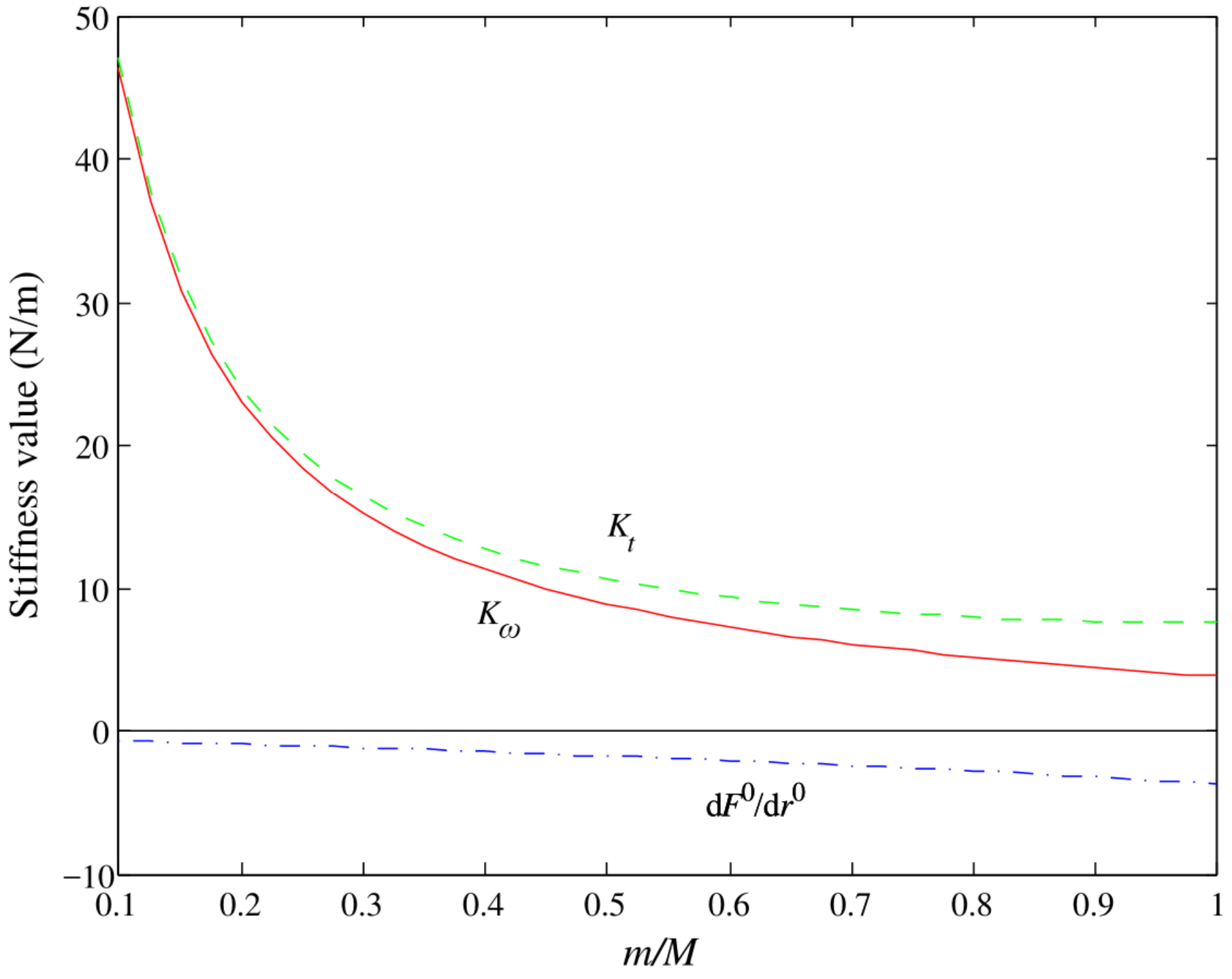

**Fig. 12.** The distributions of $K_\omega$, $K_t$ and $\mathrm{d}F^0/\mathrm{d}r^0$ on the 201mm-length spring at $\omega = 13.61\,\mathrm{rad/s}$.

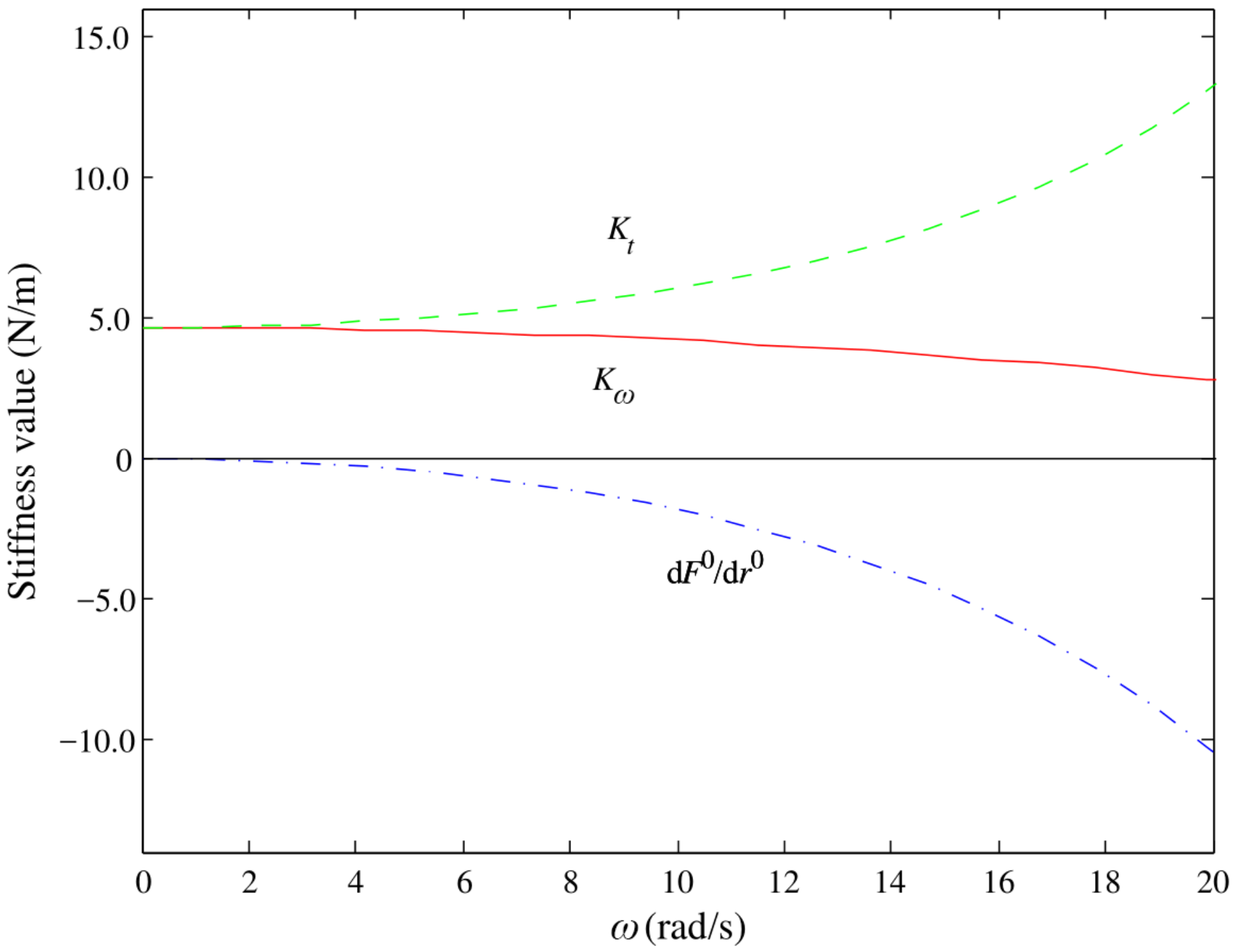

**Fig. 13.** $K_\omega$, $K_t$ and $\mathrm{d}F^0/\mathrm{d}r^0$ at the outer end of the 201mm-length spring with respect to $\omega$.

## 4. Conclusions

This paper proves both theoretically and experimentally that a heavy soft spring in steady rotation follows the 1D static Willis-form equations. This coincides with the general conclusion that the Willis-form equations are more accurate than classical linear elastic equations for inhomogeneous media with pre-stresses. Thus, it gives a clear reasoning from the original papers of Willis (1997) and Milton et al. (2006), to Xiang (2014), Xiang and Yao (2016). Another interesting observation is that the Willis-form equations can simultaneously explain the stress-stiffening and the spin-softening effect of a rotational spring with an added end mass. All these findings will be helpful to the further applications of the Willis-form equations.

## Acknowledgment

This work was supported by the National Science Foundation of China [grant numbers 11672144, 11272168].